\documentclass[twocolumn,
aps,
pre, 
notitlepage,
superscriptaddress
]{revtex4-1}

\usepackage{amsmath}
\usepackage{amssymb}
\usepackage{graphicx,color}
\usepackage{mathtools}
\usepackage{tabularx}

\begin{document}
\title{Exotic phase transitions of $k$-cores in clustered networks}
\author{Uttam \surname{Bhat}}
\affiliation{Department of Physics, Boston University, Boston, MA 02215, USA}
\affiliation{Santa Fe Institute, Santa Fe, NM 87501, USA}
\author{Munik \surname{Shrestha}}
\affiliation{Department of Mathematics and Statistics, University of Vermont, Burlington, VT 05405, USA}
\author{Laurent \surname{H\'ebert-Dufresne}}
\affiliation{Santa Fe Institute, Santa Fe, NM 87501, USA}

\begin{abstract}
The giant $k$-core --- maximal connected subgraph of a network where each node has at least $k$ neighbors --- is important in the study of phase transitions and in applications of network theory.
Unlike Erd\H{o}s-R\'enyi graphs and other random networks where $k$-cores emerge discontinuously for $k\ge 3$, we show that transitive linking (or triadic closure) leads to 3-cores emerging through single or double phase transitions of both discontinuous and continuous nature. 
We also develop a $k$-core calculation that includes clustering and provides insights into how high-level connectivity emerges.
\end{abstract}
\maketitle

\section{Introduction}

The emergence of high-level or large-scale connectivity patterns on a network influences its macroscopic behaviour. It is at times essential, as it is for traffic on transportation networks or communication on the Internet, and at times detrimental, such as epidemics on social networks or cascades of failure on technological networks. One way or the other, identifying \textit{when} and \textit{how} large-scale connectivity appears or disappears is critical. This explains the explosion of models to produce phase transitions of rich and diverse nature on networks \cite{colomer2014double, d2015anomalous, baxter2016correlated}. We here discuss how single or double phase transitions of both discontinuous and continuous nature can occur solely due to the structure of a network, without ad hoc rules or complicated models. To do so, we focus on random networks with clustering and investigate the emergence of $k$-cores.

Giant $k$-cores are the largest maximal connected sub-graphs of the network where each node has at least $k$ neighbors within the subgraph. We use the word `giant' when the largest $k$-core component scales linearly with the size of the network. Network $k$-cores are important for network analysis \cite{carmi07,alvarez2005large}, diffusion of information \cite{gonzalez2011dynamics}, and the spread of diseases \cite{kitsak2010identification}. A well-known result is that the giant $k$-core of the Erd\H{o}s-R\'enyi graphs  \cite{erdos-renyi} 
emerges discontinuously  \cite{pittel1996sudden, dorogovtsev2006k} if $k\ge3$, whereas giant $1$-core (the largest connected component) and $2$-core emerge continuously. 

However, many empirical studies \cite{Watts1998,Girvan2002,Newman2006} have shown that most real-world networks are highly clustered, i.e. they have a large number of triangles and higher order motifs, which cannot be accounted for by simple random network models like Erd\H{o}s-R\'enyi (ER) networks \cite{erdHos1961strength}, block models \cite{holland1983stochastic,PhysRevE.84.066106}, or the Configuration Model \cite{molloy1995critical, newman2001random}.
We here consider the simplest way in which we can extend the ER random network model to incorporate clustering by using transitive linking or triadic closure \cite{rapoport1953spread, bianconi2014triadic}. We find that despite its simplicity, this ensemble of networks shows remarkably rich features in its $k$-core structure. In particular, we focus on the emergence of giant 3-cores (henceforth we omit the word `giant') --- and calculate their sizes in more realistic networks with clustering.

Our results challenge the typical assumption of simple and discontinuous emergence of $k$-core structure. Our conclusions therefore have important implications on resilience \cite{cohen2000resilience}, outbreaks of diseases \cite{kitsak2010identification} or social contagions \cite{dodds2004universal, campbell2013complex}, jamming \cite{schwarz2006onset} and failures of multiple dependent networks \cite{buldyrev2010catastrophic}. All of these processes can be interpreted as generalizations to higher order constraints of classic bond percolation. They thus all depend, either directly or indirectly, on the $k$-core structure and clustering of connections in a network.

\begin{figure}[b!]
\centerline{\includegraphics[width=0.4\textwidth, trim={0cm 1.5cm 0cm 2.5cm},clip]{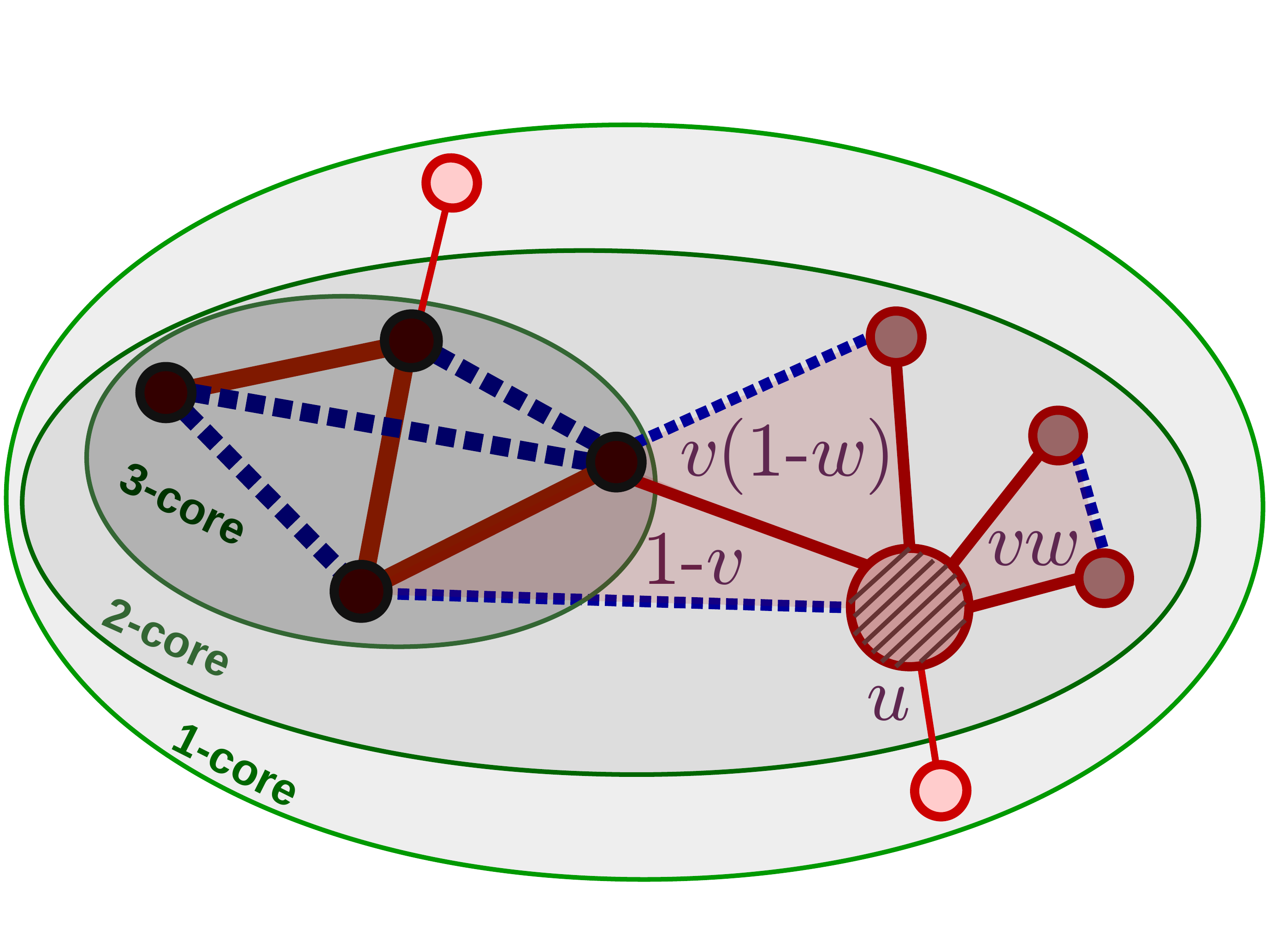}}
\caption{1-core, 2-core and 3-cores of an example network from the $(c,q)$-process. The solid (dashed) lines represent ER- (clustered-) links. The motifs around the larger hatched node are labeled by their probabilities of occurring. I.e., $u$ is the probability that a link not part of a triangle does not lead to the 3-core; $(1-v)$ is the probability that both edges of a triangle lead to the 3-core; while $v w$ is the probability that neither of the edges of the triangle lead to the 3-core.}
\label{cartoon-kore}
\end{figure}

\section{Model}
Our model, which we call the $(c,q)$-process, is a one-step extension to the classic ER random network. It has two parameters, $c$, the average degree of the ER substrate, and $q$ the probability of transitive links. We generate an ensemble of random networks with the parameters $c$ and $q$ as follows. First, draw an ER random network with average degree $c$. On this network, identify all pairs of nodes joined by a mutual neighbor (equivalently all motifs with three nodes and two links). Connect each of these pairs independently with probability $q$. Note that $q$ controls the level of clustering in the network. Henceforth we call links from the ER substrate as `ER-links' and links from the transitive linking process as `clustering' links.

A node with ER degree $k$, has $k(k-1)/2$ pairs of neighbors. Each pair is linked with probability $q$. Hence, the total number of clustering links is given by $\sum_k N e^{-c} \frac{c^k}{k!}\frac{k(k-1)}{2}q = N c^2 q/2$ yielding a total average degree of $\left\langle k\right\rangle = c + c^2 q$. A more careful, but similar calculation gives us the full degree-distribution. Each node with ER degree $k_1$, has second-neighbors distributed as the sum of $k_1$ poisson random variables with average excess degree $c$ (average number of links from a node reached by a random link, not counting the link you came from) \cite{newman2001random}. The sum of $k_1$ Poisson random variables with average $c$ is again a Poisson random variable with average $k_1 c$. Each of these second-neighbors is linked to the node with probability $q$, yielding a Poisson distributed number of clustering links with average $k_1 c q$. Thus the total degree distribution is given by,
\begin{equation}
\label{degreedist}
n_k = \sum_{k_1 + k_2 = k} \left(\frac{e^{-c} c^{k_1}}{k_1!}\right)\left( \frac{e^{-c k_1 q} (c k_1 q)^{k_2}}{k_2!}\right)
\end{equation} 
where the individual terms within the summation are the joint probability distribution of a node having $k_1$ ER links and $k_2$ clustering links.

\section{Results}
  \begin{figure}[h]
  \centerline{\includegraphics[width=0.4\textwidth, trim={0.2cm 0.2cm 0.5cm 0.4cm},clip]{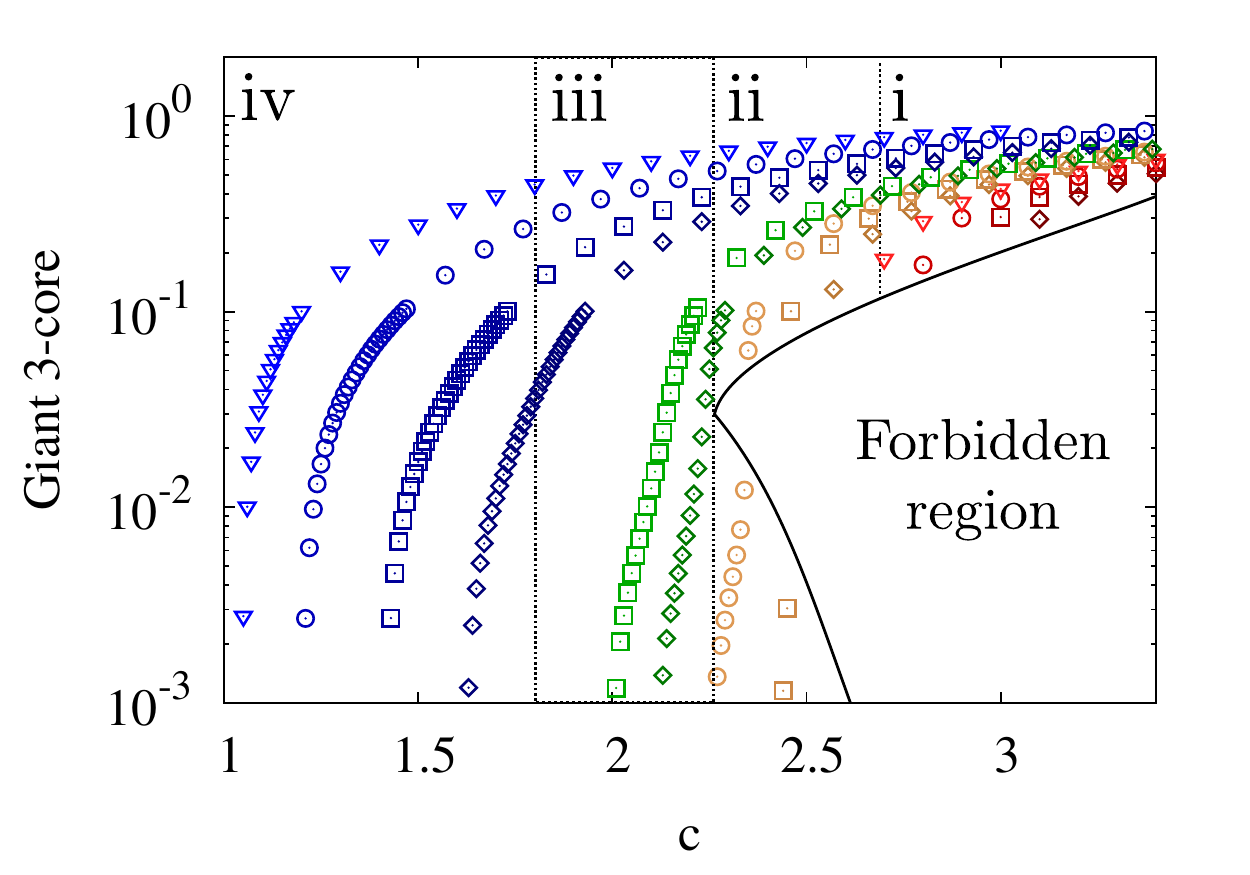}}
  \caption{(top) Plot showing the 3-core for representative values of $q$. The values of $q$ from right to left in different regimes are (i) 0.02, 0.03, 0.04, 0.05, (ii) 0.06, 0.07, 0.08, (iii) 0.09, 0.1, (iv) 0.15, 0.2, 0.3 and 0.5. On the extreme right we see the classic discontinuous emergence (regime i), whereas the other three regimes show very different behavior. The forbidden region corresponds to impossible 3-core sizes for a given $c$; its solid black borders are drawn to highlight this region.}
  \label{3core-vs-c}
 \end{figure}
 
 In this paper, we are only interested in the $k$-core structure of this model. Note that the transitive linking process does not change the size of the 1-core as it only links nodes that are already in the same component. However, the $(c,q)$-process can upgrade nodes in the 1-core component to the 2-core by connecting the nodes of degree-1 to nodes of higher degree, but since in the ER substrate both the giant 1-core and 2-core emerge at exactly $c=1$, we expect no difference after the transitive linking process. The emergence of the 3-core as a function of $c$ exhibits four distinct regimes depending on $q$,
 \begin{enumerate}
 \item[i.] a regime of sudden discontinuous appearance of a 3-core when $q$ is lesser than some bound $q^*\approx 0.06$ (we notate $q\lesssim0.06$),
 \item[ii.] a double phase transition regime consisting of a continuous appearance of a 3-core followed by a discontinuous jump in size when $0.06\lesssim q\lesssim0.08$.
 \item[iii.] another double phase transition regime consisting of two continuous transitions when $0.08\lesssim q\lesssim0.15$,
 \item[iv.] a single sudden but continuous appearance of a 3-core for $q\gtrsim0.15$.
 \end{enumerate}

\begin{figure}[h]
  \centerline{\includegraphics[width=0.45\textwidth, trim={0cm 0cm 0cm 0cm},clip]{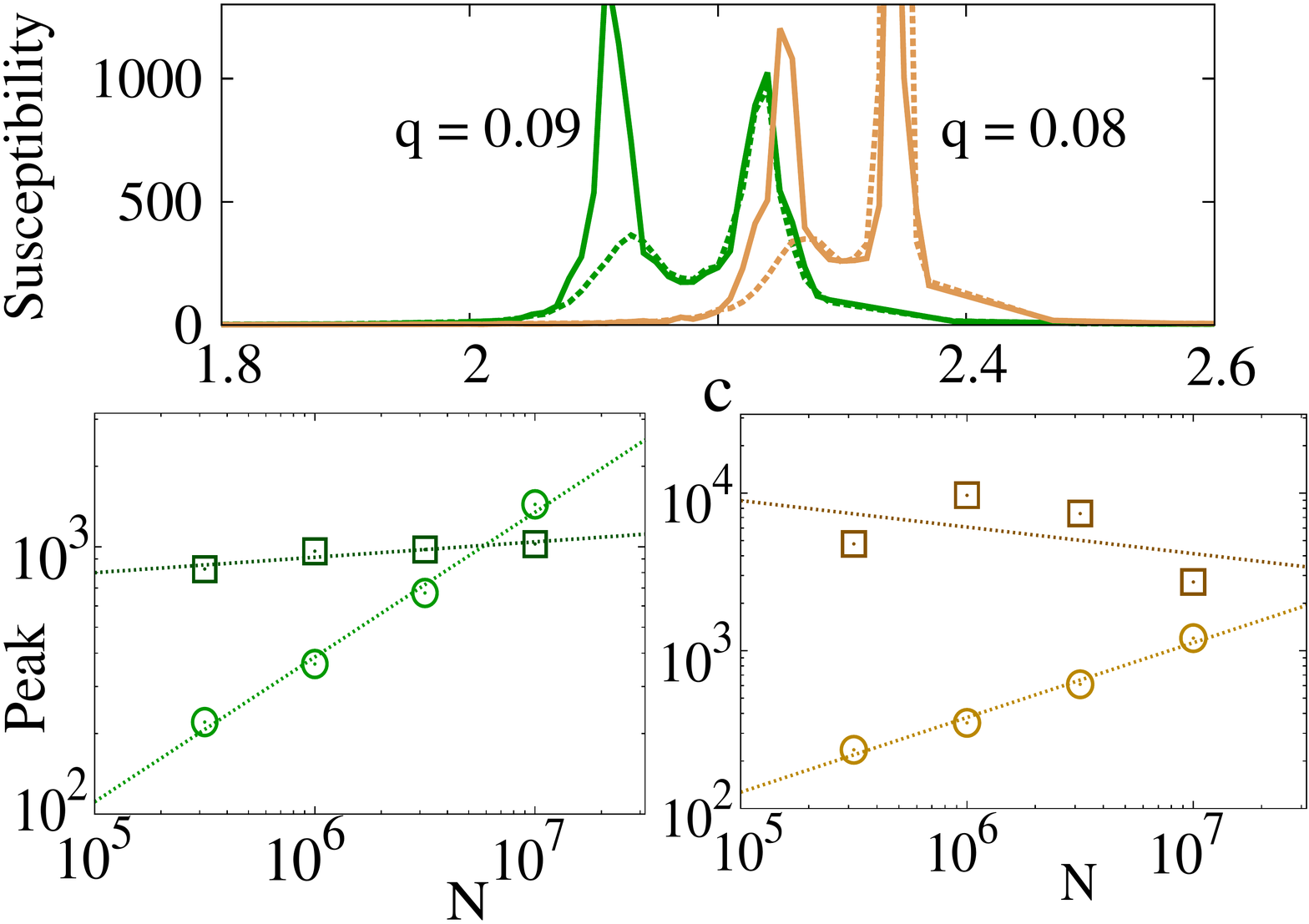}}
  \caption{(top) The susceptibility  of our simulations for representative values of $q$ in regime ii and iii highlighting a double phase transition. Dashed and solid lines are obtained on networks of size $10^6$ ($10^3$ realizations) and $10^7$ ($10^2$ realizations) respectively. (bottom) The heights of the peaks of susceptibility for networks of size $N=10^{5.5}$ through $N=10^7$. The number of realizations $= 10^9/N$. The circles (squares) show the heights of the left (right) peaks. The fits to the left peaks show a slope of $0.54\pm0.04$ for $q=0.09$ and $0.47\pm0.04$ for $q=0.08$. The fits to the right peak show slopes that are statistically indistinguishable from zero. }
  \label{susceptibility-peak}
 \end{figure}

We gather more evidence for the existence of regime ii and iii as follows. We first study the susceptibility (variance over average as defined in \cite{colomer2014double}) of the giant 3-core seen in the simulations. We see that the susceptibility shows two peaks as a function of $c$, for each value of q in these regimes. The scaling of the first peak with system size confirms that this is indeed a phase transition and not a finite size effect. Note also that, in both regimes, the height diverges for the first peak, suggesting a second order transition, and saturates for the second peak, suggesting a first order transition (Fig.~\ref{susceptibility-peak}). However, considering that we do not see a discontinuous gap at the second transition of regime iii, the fact that the order parameter is already at a non-zero value might explain the saturation of susceptibility.

\begin{figure}[h]
  \centerline{\includegraphics[width=0.45\textwidth, trim={0cm 0cm 0cm 0cm},clip]{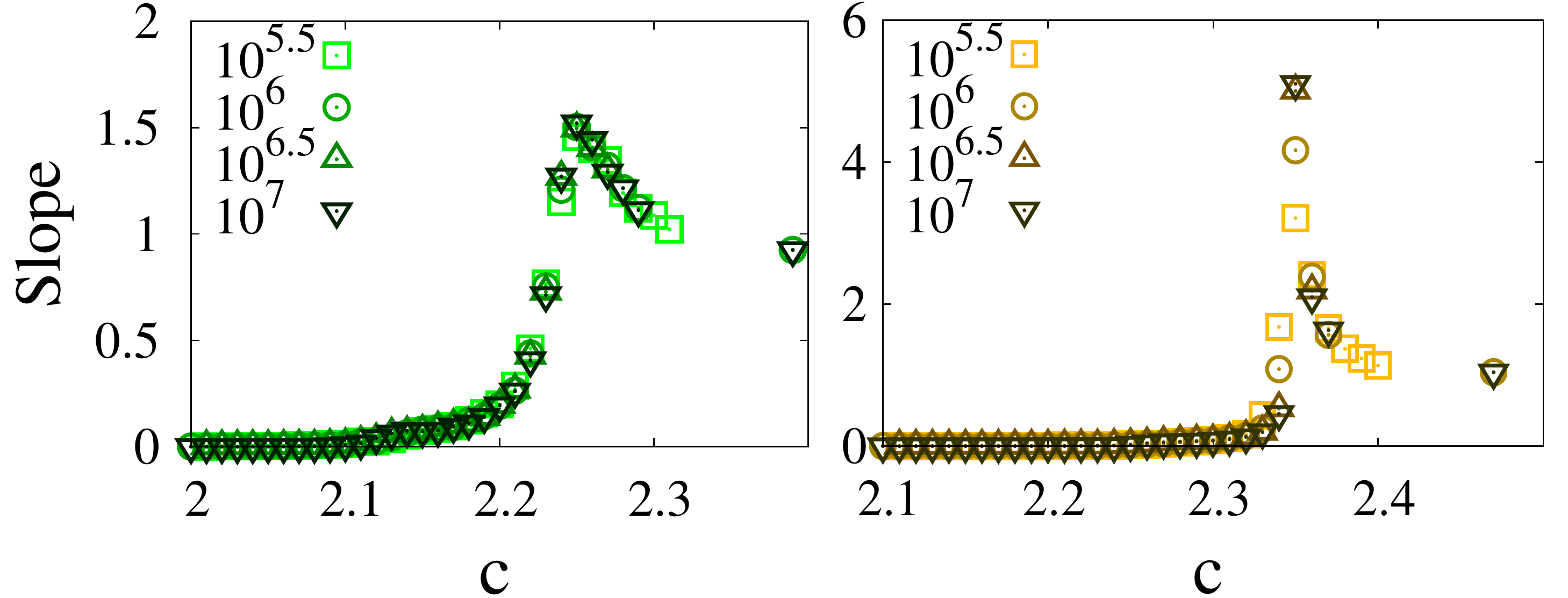}}
  \caption{Panel to the left (right) showing the jump (divergence) in the slope of the giant 3-core size as a function of $c$ for $q = 0.09\, (0.08)$.}
  \label{slope}
 \end{figure}

We further look at the slopes of the giant 3-core for values of $q$ in these regimes, and values of $c$ close to the second peak in Fig.~\ref{slope}. The slopes however, suggest that the second transition is first order in regime ii since the slope diverges, but second order in regime iii the slope exhibits a jump discontinuity, but does not diverge.
 
We believe the reason for the existence of double transition is that the 3-core first emerges through the triangles only and then eventually spread through regular links too. 

To understand the effect of $q$ on the size of the 3-core, we develop the following analytical approximations: We calculate the 3-core sizes in random networks with the same degree distribution as our $(c,q)$-process and in random networks with the same joint degree-triangle distribution as our $(c,q)$-process. These calculations lead us to conclude that the degree-distribution and clustering by themselves are not a sufficient to explain the regimes of double transition. This also highlights the role of higher order motifs in the $(c,q)$-process.\\

\subsection{Configuration model\label{CM}}
 \begin{figure}[h]
 \centerline{\includegraphics[width=0.4\textwidth, trim={0.2cm 0.2cm 0.5cm 0.4cm},clip]{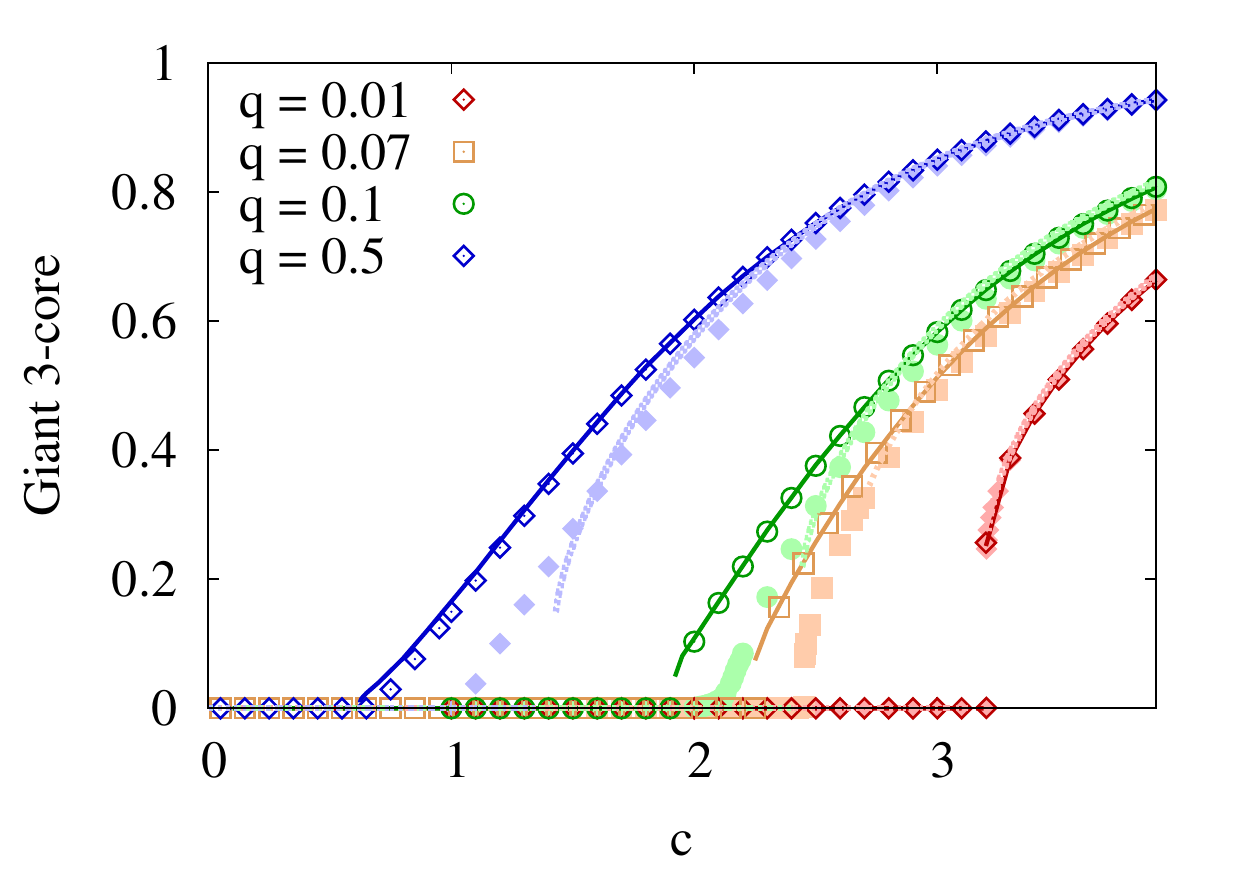}}
 \caption{Comparison between simulations of the $(c,q)$-process (solid light markers) and analytical predictions using the configuration model (dashed light lines) and the clustered configuration model (solid dark lines). We also compare the clustered configuration model to the rewired $(c,q)$-networks preserving degree-triangle distributions (open dark symbols)}
 \label{clustered-and-config-model}
 \end{figure}

 The Configuration Model defines the ensemble of all random networks constrained by a given degree distribution. We can find the size of the 3-core for this ensemble through a set of self-consistent equations \cite{goltsev2006k, PhysRevE.89.022805}. The probability $u$, that a randomly chosen link does \emph{not} lead to the 3-core is given by,
 \begin{equation}
 u = \frac{1}{G^{(1)}(1)}\sum_{a=0}^1 \frac{(1-u)^a}{a!}G^{(a+1)}(u)
 \label{uconfig}
 \end{equation}
where $G(z)$ is the generating function for the degree distribution \cite{newman2001random} and $G^{(j)}(z)$ is its $j$-th derivative.
From Eq.~\eqref{degreedist}, we can find that $G(z)$ is given by,
\begin{equation}
G(z) = \exp\left[c(z e^{c q (z-1)}-1)\right] \; .
\end{equation}
Equation \eqref{uconfig} means that for the link to not lead to the 3-core (left-hand side), the node it leads to must have at most one excess neighbor in the 3-core (right-hand side). 
We then calculate the fraction $S$ of nodes in the 3-core by summing the probability that a randomly chosen node has at least three links leading to the $3$-core (or one minus the probability of having less than three):
  \begin{equation}
  S = 1-\sum_{a=0}^2 \frac{(1-u)^a}{a!} G^{(a)}(u) \; .
  \end{equation}

This approach always leads to discontinuous transitions (Fig.~\ref{clustered-and-config-model}), which shows that the degree distribution alone is not sufficient to explain the existence of the three other regimes. Hence, we develop the following approach to incorporate clustering.

\subsection{Clustered configuration model\label{CCM}}

 The Clustered Configuration Model defines the ensemble of all random networks constrained by a given joint degree-triangle distribution $p_{k t}$. We wish to calculate the size of the $3$-core given this probability $p_{k t}$ of a randomly chosen node having $k$ links and $t$ triangles. This ensemble of networks was introduced to study the impact of clustering on bond percolation \cite{newman2009random}. In networks without clustering, all links were considered independent (as they are indeed independent in a fully random network). However, this assumption fails when a network contains triangles and we must therefore explicitly account for triangles through additional self-consistent equations. 

Let $q_{k t} = k\,p_{k t}/\left\langle k\right\rangle$ and $r_{k t} = t\,p_{k t}/\left\langle t \right\rangle$ be the excess degree-triangle distribution when following a link or a triangle, respectively \cite{newman2009random}. We can write self-consistent equations for the probability $u$ of following a link to a node \textit{not} in the 3-core, the probability $v$ that following a triangle does \textit{not} lead to two nodes in the 3-core and the probability $vw$ that following a triangle leads to exactly \textit{zero} node in the 3-core. Figure \ref{cartoon-kore} gives a visual representation of these probabilities and motifs. Mathematically, we write

\begin{widetext}
\begin{tabularx}{0.98\textwidth}{@{}XX@{}}
  \begin{equation}
  u   \; = \;  \sum_{k,t} q_{k t} \left[\sum_{a+c\le1} \mathcal{B}_a^k(1-u) v^t \mathcal{B}_c^t(1-w)\right]
    \label{eqn:u}
  \end{equation} &
  \begin{equation}
  1-v   \; = \; \left[1-\sum_{k,t}r_{k t} u^k (w v)^t\right]^2
    \label{eqn:v}
  \end{equation}
\end{tabularx}
\begin{align}
 v w \; &= \; \left(\sum_{k,t} r_{k t} u^k (w v)^t\right)\left[\sum_{k,t}r_{k t}\sum_{a+c\le1}\mathcal{B}_a^k(1-u) v^t \mathcal{B}_c^t(1-w) + \sum_{k,t}r_{k t}\sum_{a\le 1}\mathcal{B}_a^k(1-u) v^t \mathcal{B}_{1-a}^t(1-w)\right]
\label{eqn:w}
\end{align}

where $\mathcal{B}_i^n(z) := \binom{n}{i}z^i(1-z)^{n-i}$. Equation (\ref{eqn:u}) is equivalent to Eq. (\ref{uconfig}), a regular link will not lead to the 3-core if the node it reaches has at \textit{most} one link leading to the 3-core. This potential link can either be a regular link (subscript $a$ for the binomial term $\mathcal{B}_a^k(1-u)$) or be one link of a triangle (subscript $c$ for the binomial term $\mathcal{B}_c^t(1-w)$), but no triangles can contribute more than a single link (the term $v^t$). 
Equation (\ref{eqn:v}) corresponds to following a triangle from one node and finding that both links lead to the 3-core, hence both nodes reached must have at \textit{least} one other link leading to the 3-core. We thus write $1-v$ has the square of one minus the probability of having zero link leading to the 3-core, since the other links of these two nodes are independent.
Equation (\ref{eqn:w}) is more involved but similar. When following a triangle, if neither of the two links lead to the 3-core, both nodes must have at \textit{most} one link leading to the 3-core, but they can not both have one as there is also a link between the two.

The size of the 3-core, $S$, is then given by,
\begin{equation}
S = 1 - \sum_{k,t} p_{k t} \sum_{b\le 1}\;\sum_{a\le 2}\;\sum_{c\le2-a-2 b}\mathcal{B}_a^k(1-u) \mathcal{B}_b^t(1-v) \mathcal{B}_c^{t-b}(1-w)
\label{Suvw}
\end{equation}
\end{widetext}
 
 We evaluate this formula on our $(c,q)$-process as follows. We numerically obtain $p_{k t}$ from generated networks and use this in Eq.~\eqref{eqn:u}~--~\eqref{Suvw} to obtain Fig.~\ref{clustered-and-config-model}. We compare this to the 3-core sizes obtained from rewired networks preserving $p_{k t}$ and from our $(c,q)$-process simulations. We see that there is remarkable agreement except at the onset of the transition. It is important to note that the rewired networks show a systematic quantitative difference as compared to our model. This is due to the high density of adjacent triangles that share a link. Under a configuration model rewiring, these adjacent triangles are very unlikely, which means that the rewired networks have a higher average degree from the same $p_{k t}$ joint distribution. Every triangle contributes two links in the clustered configuration model whereas, in the $(c,q)$-process, we see many pairs of adjacent triangles that account for only three links as they overlap.
 
 More importantly, the clustered configuration model captures how the size of the discontinuous jump vanishes as clustering increases. However, the transition is still discontinuous and we do not see regimes of double phase transitions. Surprisingly, we can thus conclude that clustering alone does not explain regimes ii through iv; they are most likely caused the high number of adjacent triangles in the $(c,q)$-process which create many non-trivial motifs.
 
 
 \section{Discussion}
 Our model is a simple extension of the classic ER network to include clustering through triadic closure. The degree distribution still has a Gaussian tail, but remarkably we find four distinct regimes shaping how 3-cores can emerge: through a single discontinuous phase transition ($q\lesssim0.06$), through a double transition of hybrid nature ($0.06\lesssim q\lesssim0.08$), through two consecutive continuous transitions ($0.08\lesssim q\lesssim0.15$), or through a single continuous phase transition ($q\gtrsim0.15$).
 
 The first and last regime were previously obtained in heterogeneous $k$-core percolation where different nodes have different thresholds \cite{cellai2011tricritical}. Similarly, the regime of hybrid phase transitions ($0.06\lesssim q\lesssim0.08$) was observed in the completely different context of percolation on interdependent networks \cite{allard2015general}. 
These analogies might hold the key to understanding the mechanisms at play behind these different regimes: they all use a mixture of nodes with different sensitivity to ``criticality''. 
Criticality here means whether a given node is in the giant component or not (either a giant percolating component, or a giant $k$-core).
Interdependent networks use a mixture of nodes that are either independent or dependent on a given node in a different networks.
Heterogeneous $k$-core percolation uses a mixture of nodes with different thresholds; such that the giant component is a mixture of, for example, the 2-core and the 3-core.
However we show that realistic networks can have very similar properties even using the simplest definition of $k$-cores where all nodes obey the same rules. 
In this context, these different sensitivities are structural (e.g. is a node part of an important motif) rather than \textit{ad hoc} conditions.
 
  Finally, we showed a simple way to extend the configuration model to incorporate clustering in calculations for the size of the 3-core. This way of evaluating motifs can be extended straightforwardly to higher $k$-cores, and even to higher order motifs by introducing more self-consistent equations. We believe that this new analytical $k$-core calculation will provide useful estimates for the aforementioned applications as well as for the emergence and growth of $k$-core structure in real-world networks.

\begin{acknowledgments}
The authors thank Cris Moore for helpful discussions, as well as Sidney Redner and Antoine Allard for comments on the manuscript. This work has been supported by the Santa Fe Institute, by the University of Vermont, by the James S. McDonnell Foundation Postdoctoral Fellowship (LHD), by Grant DMR-1623243 from the National Science Foundation (UB), and Grant No.\ 2012145 from the United States Israel Binational Science Foundation (UB).
\end{acknowledgments}

\end{document}